\documentclass[sigconf,authorversion]{acmart}

\setcopyright{none}

\begin{document}
\title{Inaudible Voice Commands}

\author{
Liwei Song, Prateek Mittal\\
\href{mailto:liweis@princeton.edu}{liweis@princeton.edu}, \href{mailto:pmittal@princeton.edu}{pmittal@princeton.edu}\\
Department of Electrical Engineering, Princeton University
}

\begin{abstract}
Voice assistants like Siri enable us to control IoT devices conveniently with voice commands, however, they also provide new attack opportunities for adversaries. Previous papers attack voice assistants with obfuscated voice commands by leveraging the gap between speech recognition system and human voice perception. The limitation is that these obfuscated commands are audible and thus conspicuous to device owners. In this paper, we propose a novel mechanism to directly attack the microphone used for sensing voice data with \emph{inaudible voice commands}. We show that the adversary can exploit the microphone's non-linearity and play well-designed \emph{inaudible ultrasounds} to cause the microphone to record normal voice commands, and thus control the victim device inconspicuously. We demonstrate via end-to-end real-world experiments that our inaudible voice commands can attack an Android phone and an Amazon Echo device with high success rates at a range of 2-3 meters.
\end{abstract}


\maketitle

\section{Introduction}

Voice is becoming an increasingly popular input method for humans to interact with Internet of Things (IoT) devices. With the help of microphones and speech recognition techniques, we can talk to voice assistants, such as Siri, Google Now, Cortana and Alexa for controlling smart phones, computers, wearables and other IoT devices. Despite their ease of use, these voice assistants also provide adversaries new attack opportunities to access IoT devices with voice command injections.

Previous studies about voice command injections target the speech recognition procedure. Vaidya et al.~\cite{Vaidya:voice2015} design garbled audio signals to control voice assistants without knowing the speech recognition system. Their approach obfuscates normal voice commands by modifying some acoustic features so that they are not human-understandable, but can still be recognized by victim devices. Carlini et al.~\cite{Carlini:voice2016} improve this black-box approach with more realistic settings and propose a more powerful white-box attack method based on knowledge of speech recognition procedure. Although not human-recognizable, these obfuscated voice commands are still conspicuous, as device owners can still hear the obfuscated sounds and become suspicious.

\begin{figure}[!ht]
\centering
\includegraphics[width=3.3in]{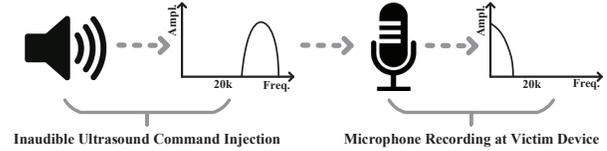}
\caption{The attack scenario for inaudible voice commands.}\label{fig:nonlinear}
\end{figure}

In contrast, we propose a novel \emph{inaudible} attack method by targeting the microphone used for voice sensing by the victim device. Due to the inherent non-linearity of the microphone, its output signal contains ``new'' frequencies other than input signal's spectrum. These ``new'' frequencies are not just integer multiples of original frequencies, but also the sum and difference of original input frequencies. Based on this security flaw, our attack scenario is shown in Fig. \ref{fig:nonlinear}. The adversary plays an ultrasound signal with spectrum above $20kHz$, which is inaudible to humans. Then the victim device's microphone processes this input, but suffers from non-linearity, causing the introduction of new frequencies in the audible spectrum. With careful design of the original ultrasound, these new audible frequencies recorded by the microphone are interpreted as actionable commands by voice assistant software.

In this paper, we put forward a detailed attack algorithm to obtain inaudible voice commands and perform end-to-end real-world experiments for validation. Our results show that the proposed inaudible voice commands can attack an Android phone with $100\%$ success at a distance of 3 meters, and an Amazon Echo device with $80\%$ success at a distance of 2 meters.

\section{Related Work}
Recently, a few papers have proposed attacks against data-collecting sensors. Son et al.~\cite{Son:drone2015} show that intentional resonant sounds can disrupt the MEMS gyroscopes and cause drones to crash. Furthermore, by leveraging the circuit imperfections, Trippel et al.~\cite{Trippel:WALNUT2017} achieve control of the outputs of MEMS accelerometers with resonant acoustic injections. Different from these approaches, we consider the microphone's non-linearity, so we do not need to find the resonant frequency. Instead, we need to carefully design ultrasounds that are interpreted by microphones as normal voice commands.

Roy et al.~\cite{Roy:backdoor2017} conduct a similar work, where the non-linearity of the microphone is exploited to realize inaudible acoustic data communications and jamming of spying microphones. However, their data communication method needs additional decoding procedures after the receiving microphone, and their jamming method injects strong \emph{random} noises to spying microphones. In contrast, we consider a completely different scenario, where the target microphone needs no modification and its outputs have to be interpreted as target voice commands.

\section{Ultrasound Injection Attacks}

In our attack scenario, the goal is to obtain well-designed ultrasounds which are inaudible when played but can be recorded similarly to normal commands at microphones. The victim can be any common IoT device with an off-the-shelf microphone, and it does not need any modification, except adopting the always-on mode to continuously listen for voice input, which has been used in many IoT devices such as Amazon Echo. To perform an attack, the adversary only needs to be physically proximate to the target and have the control of a speaker to play ultrasound, which can be achieved by either bringing an inconspicuous speaker close to the target or using a position-fixed speaker to attack nearby devices.

\subsection{Non-Linearity Insight}

\begin{figure}[!ht]
\centering
\includegraphics[width=3.2in]{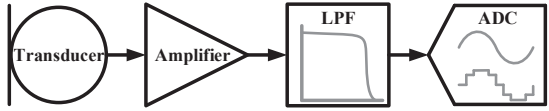}
\caption{Typical diagram of a microphone.}\label{fig:mic}
\end{figure}

As shown in Fig. \ref{fig:mic}, a typical microphone consists of four modules. The transducer generates voltage variation proportional to the sound pressure, which passes through the amplifier for signal enlargement. The low-pass filter (LPF) is then adopted to filter out high frequency components. Finally, the analog to digital converter (ADC) is used for digitalization and quantization. Since the audible sound frequency ranges from $20 Hz$ to $20 kHz$, a typical sampling rate for ADC is $48 kHz$ or $44.1 kHz$, and the filter's cut-off frequency is usually set about $20 kHz$ .

To obtain a good-quality sound recording, the transducer and the amplifier should be fabricated as linear as possible. However, they still exhibit non-linear phenomena in practice. Assume the input sound signal is $S_{in}$, the output signal after amplifier $S_{out}$ can be expressed as
\begin{equation}\label{eq:nonlinear}
S_{out} =  \sum_{i=1}^{\infty}G_{i} S_{in}^{i} = G_{1} S_{in} + G_{2} S_{in}^{2} + G_{3} S_{in}^{3} + \cdots,\\
\end{equation}
where $G_{1} S_{in}$ is the linear term and dominates for input sound in normal range. The other terms reflect the non-linearity and have an impact for a large input amplitude, usually the third and higher order terms are relatively weak compared to the second-order term.

The non-linearity introduces both \emph{harmonic distortion and intermodulation distortion} to the output signal. Suppose the input signal is sum of two tones with frequencies $f_{1}$ and $f_{2}$, i.e., $S_{in} = \cos(2 \pi f_{1}t) + \cos(2 \pi f_{2}t)$, the output due to the second-order term is expressed as
\begin{equation}\label{eq:second}
\begin{aligned}
G_{2} S_{in}^2  =  & G_{2} +  \frac{G_{2}}{2} \left(\cos\left(2 \pi \left(2f_{1}\right) t \right) + \cos \left(2 \pi \left(2f_{2}\right) t \right) \right) \\
& + G_{2} \left( \cos\left( 2 \pi \left( f_{1}\!+\!f_{2}\right) t \right) + \cos\left( 2 \pi \left( f_{1}\!-\!f_{2}\right) t \right) \right),\\
\end{aligned}
\end{equation}
which includes both harmonic frequencies $2f_1, 2f_2$ and intermodulation frequencies $f_{1} \pm f_{2}$.

Our attack intuition is to exploit the intermodulation to obtain normal voice frequencies from the processing of ultrasound frequencies. For example, if we play an ultrasound with two frequencies $25 kHz$ and $30 kHz$, the listening microphone will record the signal with the frequency of $30kHz - 25kHz = 5kHz$, while other frequencies are filtered out by the LPF.

\subsection{Attack Algorithm}

Now, we present how this non-linearity can be leveraged to design our attack ultrasound signals. Assume the signal of normal voice command, such as ``OK Google'', is $S_{normal}$. Our attack algorithm contains the following steps.

\textbf{Low-Pass Filtering}

First we adopt a low-pass filter on the normal signal, with the cut-off frequency as $8 kHz$ to remove high frequency components. Human speech is mainly concentrated on low frequency range, and many speech recognition systems, such as CMU Sphinx, only keep spectrum below $8 kHz$. Therefore, the filtering step can allow us to adopt a lower carrier frequency for modulation, while still preserving enough data of the original signal. Denote the filtered signal as $S_{filter}$.

\textbf{Upsampling}

Usually, the normal voice command $S_{normal}$ is recorded with sampling rate of $48 kHz$ (or $44.1 kHz$), the same as $S_{filter}$. This sampling rate only supports generating ultrasound with frequency ranging from $20 kHz$ to $24 kHz$ (or $22.05 kHz$), which is not enough. To shift the whole spectrum of $S_{filter}$ into inaudible frequency range, the maximum ultrasound frequency should be no less than $28kHz$. Thus, we derive an upsampled signal $S_{up}$ with higher sampling rate.

\textbf{Ultrasound Modulation}

In this step, we need to shift the spectrum of $S_{up}$ into high frequency range to be inaudible. Here, we adopt amplitude modulation for spectrum shifting. Assuming the carrier frequency is $f_c$, the modulation can be expressed as
\vspace{-2mm}
\begin{equation}\label{eq:dsb}
S_{modu} = n_1 S_{up} \cos(2\pi f_c t),\\
\end{equation}
where $n_1$ is the normalized coefficient. The resulting modulated signal contains two sidebands around the carrier frequency, ranging from $f_c - 8kHz$ to $f_c + 8kHz$. Therefore, $f_c$ should be at least $28 kHz$ to be inaudible.

\textbf{Carrier Wave Addition}

Modulating the voice spectrum into inaudible frequency range is not enough, they have to be translated back to normal voice frequency range at the microphone for successful attacks. Without modifying the microphone, we can leverage its non-linear phenomenon to achieve demodulation by adding a suitable carrier wave, and the final attack ultrasound can be expressed as
\begin{equation}\label{eq:am_attack}
S_{attack} = n_2 (S_{modu} + \cos (2 \pi f_c t)),\\
\end{equation}
where $n_2$ is used for signal normalization.

The above steps illustrate the entire process of obtaining an attack ultrasound. This well-designed inaudible signal $S_{attack}$, when played by the attacker, can successfully inject a voice signal similar to $S_{normal}$ at the target microphone and therefore control the victim device inconspicuously.

\section{Evaluation}

We perform real-world experiments to evaluate our proposed inaudible voice commands. All of the following tests are performed in a closed meeting room measuring approximately 6.5 meters by 4 meters, 2.5 meters tall. To play the attack ultrasound signals, we first use a text-to-speech application to obtain the normal voice commands and follow the described attack algorithm with $192kHz$ upsampling rate and $30kHz$ carrier frequency to get attack signals in our laptop. Then a commodity audio amplifier \cite{rs202} is connected for power amplification, and the amplified signals are provided to a tweeter speaker \cite{ft17h}. A video demo of the attack is available at \url{https://youtu.be/wF-DuVkQNQQ}.

\subsection{Attack Demonstration}

We first validate the feasibility of our inaudible voice commands: the normal voice command is ``OK Google, take a picture'', and a Nexus 5X running Android 7.1.2 is placed 2 meters away from the speaker for recording.

\begin{figure}[!ht]
\centering
\includegraphics[width=3.3in,height=2in]{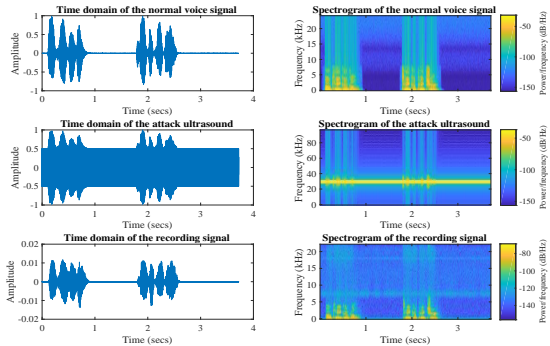}
\caption{Time plots and spectrograms for the normal voice, the attack ultrasound and the recording signal.}\label{fig:normal_record}
\end{figure}

Fig. \ref{fig:normal_record} presents the normal voice command, the attack ultrasound and the recording sound in both time domain and frequency domain. We can see that the spectrum of attack ultrasound is above $20kHz$, and after processing this ultrasound, the microphone's recording sound is quite similar to the normal voice. When playing the attack ultrasound, the phone is successfully activated and opens the camera.

\subsection{Attack Performance}

We further examine our ultrasound attack range for two devices: an Android phone and an Amazon Echo, where we try to spoof voice commands ``OK Google, turn on airplane mode'', and ``Alexa, add milk to my shopping list'', respectively. The following table shows the relationship between the attack range and the speaker's input power. We can see that the attack range is positively correlated to the speaker's power. The attack range of our approach is less for Amazon Echo compared to the Android phone, since its microphone is plastic covered.

\begin{table}[!ht]
\renewcommand{\arraystretch}{1.3}
\caption{The relationship between our attack range and the speaker's input power.} \label{tab:range}
\vspace{-3mm}
\centering
\begin{tabular}{|l|c|c|c|c|c|}
\hline Input Power ($Watt$) & $9.2$ & $11.8$ & $14.8$ & $18.7$ & $23.7$\\
\hline Range (Phone, $cm$) & $222$ & $255$ & $277$ & $313$ & $354$\\
\hline Range (Echo, $cm$) & $145$ & $168$ & $187$ & $213$ & $239$\\
\hline
\end{tabular}
\end{table}

We also check the attack accuracy by setting input power as $18.7W$ and placing phone and Echo $3m$ and $2m$ away, respectively. For each device, we repeat the corresponding inaudible voice command every 10 seconds for 50 times. The attack success rates are $100\%(50/50)$ for the Android phone and $80\%(40/50)$ for the Amazon Echo.

\section{Conclusion}

Based on the inherent non-linear properties of microphones, we propose a novel attack method by transmitting well-design ultrasounds to control common voice assistants, like Siri, Google Now, and Alexa. By taking advantage of intermodulation distortion and amplitude modulation, our attack voice commands are \emph{inaudible} and achieve high success rates on an Android phone more than three meters away and on an Amazon Echo device more than two meters away.

\bibliographystyle{ACM-Reference-Format}

\end{document}